\documentclass[twocolumn,aps, pra]{revtex4}
\usepackage[latin9]{inputenc}
\usepackage{amsmath}
\usepackage{amssymb}
\usepackage{graphicx}

\makeatletter
\@ifundefined{textcolor}{}
{%
 \definecolor{BLACK}{gray}{0}
 \definecolor{WHITE}{gray}{1}
 \definecolor{RED}{rgb}{1,0,0}
 \definecolor{GREEN}{rgb}{0,1,0}
 \definecolor{BLUE}{rgb}{0,0,1}
 \definecolor{CYAN}{cmyk}{1,0,0,0}
 \definecolor{MAGENTA}{cmyk}{0,1,0,0}
 \definecolor{YELLOW}{cmyk}{0,0,1,0}
}

\usepackage{dcolumn}
\usepackage{bm}
\usepackage{array}

\makeatother

\begin{document}

\title{Polaron in a non-abelian Aubry-André-Harper model with \textit{p}-wave
superfluidity}

\author{Xiao-Dong Bai,$^{1,2,3}$ Jia Wang,$^{1}$ Xia-Ji Liu,$^{1}$ Jun
Xiong$^{2}$, Fu-Guo Deng,$^{2,3}$ and Hui Hu$^{1}$}

\affiliation{$^{1}$Centre for Quantum and Optical Science, Swinburne University
of Technology, Melbourne 3122, Australia }

\affiliation{$^{2}$Department of Physics and Applied Optics Beijing Area Major
Laboratory, Beijing Normal University, Beijing 100875, China }

\affiliation{$^{3}$NAAM-Research Group, Department of Mathematics, Faculty of
Science, King Abdulaziz University, Jeddah 21589, Saudi Arabia}

\date{\today}
\begin{abstract}
We theoretically investigate the behavior of a mobile impurity immersed
in a one-dimensional quasi-periodic Fermi system with topological
$p$-wave superfluidity. This polaron problem is solved by using a
standard variational approach, the so-called Chevy ansatz. The polaron
states are found to be strongly affected by the strength of the quasi-disorder
and the amplitude of the $p$-wave pairing. We analyze the phase diagram
of the polaron ground state and find four phases: two extended phases,
a weakly-localized phase and a strongly-localized phase. It is remarkable
that these polaron phases are directly corresponding to the four distinct
phases experienced by the underlying background Fermi system. In particular,
the weakly-localized polaron phase corresponds to an intriguing critical
phase of the Fermi system. Therefore, the different phases of the
background system can be unambiguously probed by measuring the polaron
properties via radio-frequency spectroscopy. We also investigate the
high-lying excited polaron states at an infinite temperature and address
the possibility of studying many-body localization (MBL) of these
states. We find that the introduction of $p$-wave pairing may delocalize
the many-body localized states and make the system easier to thermalize.
Our results could be observed in current state-of-the-art cold-atom
experiments.
\end{abstract}

\maketitle

\section{Introduction}

\label{sec1}

Impurity and disorder are key ingredients of many intriguing phenomena
in quantum systems. Disorder in a non-interacting system can lead
to an unexpected phenomenon, namely Anderson localization (AL) \cite{PWAnderson1958},
which has been widely investigated and observed experimentally in
various systems, including microwaves \cite{RDalichaouch1991,AAChabanov2000},
optical waves \cite{TSchwartz2007,YLahini2008}, and matter waves
\cite{JBilly2008,GRoati2008}. Another intriguing phenomenon that
has recently received intensive interest is many-body localization
(MBL), in which a disordered, interacting many-body quantum system
fails to act as its own heat bath \cite{MRigol2008,APolkovnikov2011}
and never achieves local thermal equilibrium. As a result, MBL challenges
the very foundations of quantum statistical physics, e.g., the absence
of thermalization and a violation of the eigenstate thermalization
hypothesis (ETH) \cite{JMDeutsch1991,MSrednicki1994}. MBL also leads
to striking theoretical predictions and experimental observations
\cite{RNandkishore2015,EAltman2015}, such as the preservation of
local quantum information for a very long time \cite{APal2010} and
the slow logarithmic growth of entanglement entropy with time \cite{JHBardarson2012,MSerbyn2013,RVosk2013,FAndraschko2014}.
Remarkably, MBL was recently observed in an experiment by trapping
ultracold atoms in a one-dimensional (1D) quasi-periodic lattice \cite{MSchreiber2015},
which can be well described by the Aubry-André-Harper (AAH) model
\cite{PGHarper1955,SAubry1980,SIyer2013}.

The AAH model has been extensively applied in condensed matter systems
to investigate the transportation and AL properties of 1D quasiperiodic
systems. Based on this model, many excellent works have studied a
variety of transitions between metallic (extended), critical, and
insulating (localized) phases \cite{SOstlund1983,MKohmoto1983,DJThouless1983,JHHan1994,IChang1997,YTakada2004,FLiu2015}.
Recently, the AAH model has been extended to understand some topological
states of matter \cite{LJLang2012,WDeGottardi2013,XCai2013,JWang2016,Zeng2017,IISatija2013,YEKraus2012,SLZhu2013,FGrusdt2013,RBarnett2013,XDeng2014}.
In particular, a non-abelian extension of the AAH model that includes
$p$-wave pairing/superfluidity was used to address the interplay
between localization and non-trivial topology in a non-interacting
system \cite{WDeGottardi2013,XCai2013,JWang2016,Zeng2017}. It was
shown that, if the quasidisorder strength is large enough the system
becomes localized and topologically trivial. On the contrary, all
the states of the system are extended and topologically non-trivial,
if the quasidisorder strength is smaller than a threshold. In these
studies, the inverse participation ratio (IPR) has been applied to
characterize the phase transitions \cite{XCai2013,JWang2016,Zeng2017}.
In the thermodynamic limit $L\rightarrow\infty$, the IPR approaches
a finite value that does not depend on the size of the system $L$
in the localized phase, approaches zero as $1/L$ in the extended
phase, and decays to zero slower than $1/L$ in the intermediate critical
regime \cite{JWang2016}. In the presence of inter-particle interactions,
one may also expect to observe the MBL transition in a generalized
non-abelian AAH model, where the effect of the topologically non-trivial
$p$-wave superfluidity in MBL system can be explored.

\begin{figure}
\centering{}\includegraphics[width=0.48\textwidth]{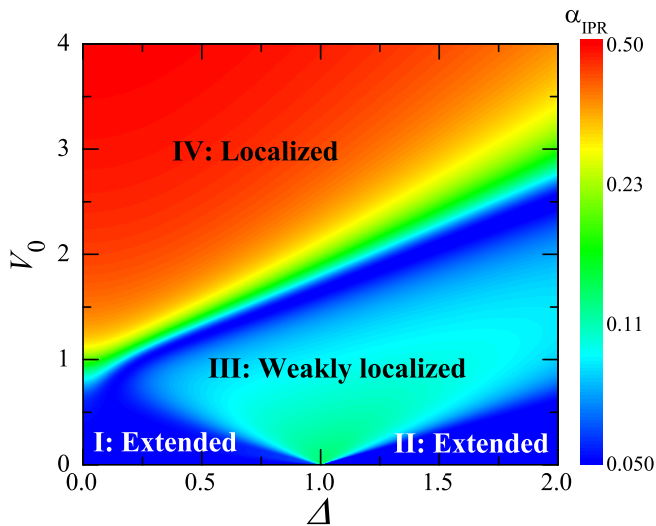}
\caption{(color online). Ground-state phase diagram of a moving impurity in
a quasi-disordered Fermi system, as functions of a $p$-wave pairing
parameter $\Delta$ and a disorder strength $V_{0}$, both of which
are measured in units of the hopping amplitude $t$. The color in
the logarithmic scale represents the value of the inverse participation
ratio $\alpha_{\texttt{IPR}}$ of the impurity wave function, from
which one may identify two extended phases (I and II) and two localized
phases (III and IV). Here, we set the length of the system $L=21$
and the offset phase of the disorder potential $\theta=\pi/(2L)$.
The interaction strengths between impurity and fermions are $U_{1}=U_{2}=2t$.}
\label{fig1} 
\end{figure}

The purpose of this work is two-fold. First, we aim to determine the
ground-state phase diagram of the generalized AAH model by introducing
a \emph{mobile} impurity that creates the so-called quasiparticle
``polaron'' as a probe \cite{FChevy2010,PMassignan2014}. In condensed
matter community, a quenched \emph{static} impurity has been widely
applied as an important local probe that characterizes the underlying
nature of the hosting quantum many-body systems \cite{AVBalatsky2006}.
For example, individual impurity has been experimentally implemented
to determine the superconducting pairing symmetry of high-temperature
superconductors \cite{EWHudson2001} and has been theoretically proposed
to probe topological superfluidity \cite{HHu2013}. Here, using a
mobile impurity (and the corresponding polaron) as a probe is similar
to a static one, but is sometimes easier to access experimentally,
particularly in the highly controllable cold-atom experiments \cite{FChevy2010,PMassignan2014},
where polaron, either Fermi polaron or Bose polaron, can be easily
created, controlled and detected. As a concrete example, we consider
a single mobile impurity immersed in a Fermi ``sea'' of two-component
fermionic atoms with \emph{p}-wave pairing as described by the generalized
non-abelian AAH model \cite{XCai2013,JWang2016}. In addition, there
is a tunable contact interaction between impurity and fermionic atoms.
We anticipate that the quasiparticle properties of the resulting polaron
should strongly depend on the underlying phases of the non-abelian
AAH model. Therefore, by measuring these quasiparticle properties
(such as the polaron energy and residue) via radio-frequency spectroscopy,
we can map out the ground-state phase diagram of the non-abelian AAH
model at zero temperature. Second, we wish to understand how the MBL
transition is affected by the \emph{$p$}-wave superfluidity in the
non-abelian AAH model. At infinite temperature, a polaron presents
one of the simplest many-body localization system \cite{HHu2016}.
The polaron may become localized with a strong enough interaction
between impurity and fermionic atoms, the phase transition thus provides
important information of the interplay between the topological superfluidity
and localization.

Our main results on the ground polaron state are briefly summarized
in Fig. \ref{fig1}. We can distinguish four different phases by the
IPR of the polaron ground state. In phases I and II, the background
AAH system is extended and the polaron IPR depends on the size of
the system $L$ and behaves like $1/L$. The IPR thus decreases to
zero when the system size is large enough. In phase IV, the system
becomes localized and the polaron IPR remains to be a finite value
that is independent of the system size. Interestingly, in phase III,
the polaron IPR exhibits a suppression with respect to the system
size $L$, but eventually approaches a finite but small value in the
thermodynamic limit $L\rightarrow\infty$. We name it as a \emph{weakly-localized}
phase and interpret it as a direct reflection of the critical phase
of the background AAH system. As a result, the large critical area
in the non-abelian AAH model due to the $p$-wave superfluidity may
be easily identified through the measurement of the IPR of the polaron.

At infinite temperature, on the other hand, the MBL transition is
also significantly affected by the existence of $p$-wave pairing.
We find that the introduction of a $p$-wave superfluidity usually
delocalizes the MBL state and makes the system easier to thermalize,
i.e., the critical disorder strength for the MBL transition increases
rapidly with increasing $p$-wave pairing (not shown in Fig. \ref{fig1}).

The remainder of this paper is organized as follows: In Sec. \ref{sec2},
we outline the model. In Sec. \ref{sec3}, we describe the details
of a variational approach to solve the single polaron problem, which
was developed by Chevy a decade ago. We also describe our diagnostics
for determining the localization of the ground polaron state and the
many-body localization at infinite temperature. In Sec. \ref{sec4}
and Sec. \ref{sec5}, we discuss the numerical results of the polaron
ground state and many-body localization, respectively. Finally, we
conclude in Sec. \ref{sec6}.

\section{The model}

\label{sec2}

Our system consists a mobile impurity immersed in a sea of noninteracting
two-component fermionic atoms loaded into a 1D non-abelian quasi-disordered
lattice. The system can be described by the model Hamiltonian, 
\begin{equation}
\mathcal{H}=\mathcal{H}_{0}+\mathcal{H}_{1},\label{totalHami}
\end{equation}
where
\begin{equation}
\mathcal{H}_{0}=\sum_{n=1}^{L}\left[\left(\mathbf{c}_{n+1}^{\dagger}\hat{T}_{1}\mathbf{c}_{n}+\textrm{H.c.}\right)+V_{n}\mathbf{c}_{n}^{\dagger}\hat{T}_{2}\mathbf{c}_{n}\right]\label{H0}
\end{equation}
is the simplest non-Abelian AAH model \cite{JWang2016} with a two-component
annihilation field operator $\mathbf{c}_{n}\equiv(c_{n,\uparrow},c_{n,\downarrow})$
and SU($N=2$) hopping matrices 
\begin{equation}
\begin{aligned}\hat{T}_{1}=t\sigma_{z}-\text{\ensuremath{i}\ensuremath{\Delta\sigma}}_{y}\end{aligned}
\label{T1}
\end{equation}
and 
\begin{equation}
\begin{aligned}\hat{T}_{2}=t\sigma_{z}.\end{aligned}
\label{T2}
\end{equation}
Here $\sigma_{z}$ and $\sigma_{y}$ are the usual 2 by 2 Pauli matrices,
and $t$ is the hopping amplitude between the nearest-neighboring
lattice sites and is set as the unit of energy (i.e., $t=1$). The
quasi-disorder lattice is characterized by 
\begin{equation}
V_{n}=2V_{0}\cos\left(2\pi n\beta+\theta\right),
\end{equation}
where $V_{0}\geq0$ is the amplitude of the quasi-disorder, $\beta$
is an irrational number that determines the quasi-periodicity, and
$\theta$ is an offset phase. If $\Delta=0$, this model $\mathcal{H}_{0}$
reduces to two identical copies of the well-known AAH Hamiltonian
(i.e., one copy for each component). In the case that $\Delta>0$,
$\mathcal{H}_{0}$ is invariant under the particle-hole transformation,
i.e., $c_{n,\uparrow}\leftrightarrow c_{n,\downarrow}^{\dagger}$.
The two spin components of the system can then be viewed as the particle
and hole components of a spinless $p$-wave superfluid and the parameter
$\Delta$ can be conveniently regarded as the $p$-wave pairing \cite{JWang2016}.
Therefore, we term the model $\mathcal{H}_{0}$ as the generalized
AAH model with $p$-wave superfluity. There are four phases in the
model $\mathcal{H}_{0}$ separated by three critical lines ($V_{0}=t+\Delta$
or $V_{0}=\left|t-\Delta\right|$) (see Fig. 1 in Ref. \cite{JWang2016}
for details). For strong quasi-disorder strength $V_{0}>t+\Delta$,
all the states of the system are localized and the system is topologically
trivial \cite{XCai2013}. For small quasi-disorder strength $V_{0}<\left|t-\Delta\right|$,
all the states are extended and the system is topologically non-trivial.
At last, for a moderate quasi-disorder strength $\left|t-\Delta\right|<V_{0}<t+\Delta$,
the three separation lines enclose a large critical area, in which
all the states of the system are multifractal. 

In Eq. (\ref{totalHami}), $\mathcal{H}_{1}$ describes the motion
of the impurity and its interactions with fermionic atoms in the lattice,
\begin{eqnarray}
\mathcal{H}_{1} & = & \sum_{n=1}^{L}\left[U_{1}c_{n,\uparrow}^{\dagger}c_{n,\uparrow}d_{n}^{\dagger}d_{n}+U_{2}c_{n,\downarrow}^{\dagger}c_{n,\downarrow}d_{n}^{\dagger}d_{n}\right]\nonumber \\
 &  & +t_{d}\left(d_{n+1}^{\dagger}d_{n}+d_{n}^{\dagger}d_{n+1}\right).\label{H1}
\end{eqnarray}
Here, $d_{n}$ is the the annihilation field operator for impurity
and $U_{1(2)}$ represents the interactions between impurity and atoms
with spin up (down). In this work, we mainly focus on the case that
the interactions are repulsive ($U_{1,2}>0$) and equal ($U_{1}=U_{2}$).
We assume that the impurity is not affected by the quasi-periodic
potential and can move freely through the lattice with a hopping amplitude
$t_{d}=t$.

Following the typical choice in the literature, for the quasi-disorder
potential, we use an irrational number $\beta=(\sqrt{5}-1)/2$, which
is the inverse of the golden mean. It can be gradually approached
by using the series of Fibonacci numbers $F_{l}$: 
\begin{equation}
\beta=\lim_{l\rightarrow\infty}\frac{F_{l-1}}{F_{l}},
\end{equation}
where $F_{l}$ is recursively defined by the relation $F_{l+1}=F_{l}+F_{l-1}$,
starting from $F_{0}=F_{1}=1$. Thus, in numerical calculations we
take the rational approximation: $\beta\simeq\beta_{l}=F_{l-1}/F_{l}$.
To minimize the possible effect of the boundary, we take the periodic
boundary condition (i.e., $c_{n+L,\sigma}=c_{n,\sigma}$ and $d_{n+L}=d_{n}$)
and assume that the length of the system $L$ is periodic with a period
$F_{l}$.

\section{Chevy's variational approach}

\label{sec3}

Inspired by the great success of Chevy ansatz in solving the polaron
problems \cite{PMassignan2014,FChevy2006}, in this section we diagonalize
the impurity Hamiltonian Eq. (\ref{totalHami}) by using the same
variational ansatz in \emph{real} space bases, within the one particle-hole
approximation.

\subsection{Chevy ansatz with one particle-hole excitation}

For the non-Abelian AAH model $\mathcal{H}_{0}$ with a two-component
field operator, expanding the wave-function in real space bases in
the form \cite{JWang2016},
\begin{equation}
\begin{aligned}|\psi\rangle=\sum_{n=1}^{L}\left[u_{n}c_{n,\uparrow}+v_{n}c_{n,\downarrow}\right]|0\rangle,\end{aligned}
\end{equation}
we can diagonalize the model Hamiltonian $\mathcal{H}_{0}$ in Eq.
(\ref{H0}) to obtain all the eigenvalues $E_{\eta}$ and the corresponding
eigenvectors \cite{JWang2016}
\begin{equation}
\begin{aligned}\psi_{\eta}=\left[u_{1,\eta},v_{1,\eta},\text{...},u_{\text{\ensuremath{n},\ensuremath{\eta}}},v_{\text{\ensuremath{n},\ensuremath{\eta}}},\text{...},u_{\text{\ensuremath{L},\ensuremath{\eta}}},v_{\text{\ensuremath{L},\ensuremath{\eta}}}\right]^{T},\end{aligned}
\end{equation}
where $n$ is the number of the lattice site, and $\eta=1,2,\text{...},2L$
is the index of the $\eta$-th single-particle state of atoms, and
$u_{n,\eta}$ and $v_{n,\eta}$ are the corresponding $\eta$-th wavefunction
at the $n$-th site. By denoting $c_{\eta}$ as the annihilation field
operator in the $\eta$-th eigenstate, we then have 
\begin{equation}
\begin{aligned}\left\{ \begin{array}{c}
c_{n,\uparrow}=\sum_{\eta=1}^{2L}u_{n,\eta}c_{\eta}\\[5pt]
c_{n,\downarrow}=\sum_{\eta=1}^{2L}v_{n,\eta}c_{\eta}
\end{array}\right..\end{aligned}
\end{equation}
Therefore, the local density of spin-up and -down atoms at the $n$-th
site is, 
\begin{equation}
\begin{aligned}\left\{ \begin{array}{c}
c_{n,\uparrow}^{\dagger}c_{n,\uparrow}=\sum_{\text{\ensuremath{\eta_{1}}\ensuremath{\eta_{2}}}}u_{n,\text{\ensuremath{\eta_{1}}}}^{*}u_{n,\text{\ensuremath{\eta_{2}}}}c_{\eta_{1}}^{\dagger}c_{\eta_{2}}\\[5pt]
c_{n,\downarrow}^{\dagger}c_{n,\downarrow}=\sum_{\text{\ensuremath{\eta_{1}}\ensuremath{\eta_{2}}}}v_{n,\text{\ensuremath{\eta_{1}}}}^{*}v_{n,\text{\ensuremath{\eta_{2}}}}c_{\eta_{1}}^{\dagger}c_{\eta_{2}}
\end{array}\right..\end{aligned}
\end{equation}

Throughout this work, we consider a Fermi sea of fermionic atoms that
are occupied up to the chemical potential $\mu\simeq0$: 
\begin{equation}
\begin{aligned}|\text{FS}\rangle=\prod_{E_{\eta}<0}c_{\eta}^{\dagger}|0\rangle,\end{aligned}
\end{equation}
which corresponds to the case of the half-filling of fermionic atoms
in the lattice, i.e., 
\begin{equation}
\left\langle c_{n,\uparrow}^{\dagger}c_{n,\uparrow}+c_{n,\downarrow}^{\dagger}c_{n,\downarrow}\right\rangle \simeq1.
\end{equation}
Here, the level index $\eta$ of the single-particle states runs from
$1$ (i.e., the ground state) to $L-1$ (i.e., $E_{\eta=L-1}<0$ but
$E_{\eta=L}>0$), and, finally, to $2L$ (i.e., the highest energy
state). Thus, we obtain the energy of fermionic atoms at zero temperature,

\begin{equation}
\begin{aligned}E_{\text{FS}}\equiv\sum_{E_{\eta}<0}E_{\eta}.\end{aligned}
\end{equation}

Following Chevy's variational approach, we take into account only
single particle-hole pair excitation. A mobile impurity may then be
described by the following approximate many-body wave function in
real space: 
\begin{equation}
\begin{aligned}|\mathcal{P}\rangle=\sum_{n}z_{n}d_{n}^{\dagger}|\text{FS}\rangle+\!\!\!\sum_{\mbox{\tiny\ensuremath{\begin{array}{c}
n,E_{\text{\ensuremath{\eta_{p}}}}>0\\
E_{\text{\ensuremath{\eta_{h}}}}<0
\end{array}}}}\!\!\!\!\alpha_{n}(\text{\ensuremath{\eta_{h}}},\text{\ensuremath{\eta_{p}}})d_{n}^{\dagger}c_{\text{\ensuremath{\eta_{p}}}}^{\dagger}c_{\text{\ensuremath{\eta_{h}}}}|\text{FS}\rangle,\end{aligned}
\end{equation}
where $z_{n}$ gives the residue of the impurity at each lattice site
$n$. The second term with amplitude $\alpha_{n}(\eta_{h},\eta_{p})$
describes the single particle-hole excitation. We note that, the site
index $n$ takes $L$ values, the level index $\eta_{h}$ (for hole
excitations) runs from $1$ to $L-1$, and $\eta_{p}$ (for particle
excitations) takes $L+1$ values. Therefore, the dimension of the
whole Hilbert space of Chevy's ansatz that we will deal with is 
\begin{equation}
\begin{aligned}D=L[1+(L-1)(L+1)]=L^{3}.\end{aligned}
\end{equation}
Thus, in our calculations, the length of the lattice and the dimension
of the Hilbert space are respectively, $L=F_{l=7}=21$ and $D_{l=7}=9,261$,
$L=F_{l=8}=34$ and $D_{l=8}=39,304$, $L=F_{l=9}=55$ and $D_{l=9}=166,375$,
and $L=F_{l=10}=89$ and $D_{l=10}=2,985,984$.

We also note that, Chevy's variational ansatz provides an excellent
description for both low-lying impurity states (i.e., attractive polarons
in the context of a mobile impurity in cold-atoms \cite{PMassignan2014})
and high-lying impurity states (i.e., repulsive polarons).

\subsection{Numerical solutions of Chevy's ansatz}

The real space bases of Chevy's ansatz within the one particle-hole
pair approximation is given by, $|i\rangle=d_{n}^{\dagger}|\text{FS}\rangle$
or $|i\rangle=d_{n}^{\dagger}c_{\eta_{p}}^{\dagger}c_{\eta_{h}}|\text{FS}\rangle$,
where the index $i$ runs from $1$ to $D$. Although the dimension
of the corresponding Hilbert space is still quite large (i.e., $D\sim10^{4}-10^{6}$),
most of the matrix elements of the interaction Hamiltonian $\mathcal{H}_{1}$
are zero. Therefore, the total Hamiltonian $\mathcal{H}=\mathcal{H}_{0}+\mathcal{H}_{1}$
can cast into a \emph{sparse} matrix that can be easily diagonalized
by standard exact diagonalization techniques \cite{HHu2016}. To be
specific, we have three kinds of matrix elements $\mathcal{H}_{ij}$
:\begin{widetext} 

\begin{equation}
\langle\text{FS}|d_{n}\mathcal{H}d_{n'}^{\dagger}|\text{FS}\rangle=\delta_{nn'}E_{\text{FS}}+U_{1}\delta_{nn'}\sum_{E_{\eta}<0}\left|u_{n,\eta}\right|{}^{2}+U_{2}\delta_{nn'}\sum_{E_{\eta}<0}\left|v_{n,\eta}\right|{}^{2}+t_{d}\delta_{n\pm1,n'},
\end{equation}

\begin{equation}
\langle\text{FS}|d_{n}\mathcal{H}d_{n'}^{\dagger}c_{\eta_{p}^{\prime}}^{\dagger}c_{\eta_{h}^{\prime}}|\text{FS}\rangle=U_{1}\delta_{nn'}u_{n,\eta_{p}^{\prime}}u_{n,\eta_{h}^{\prime}}^{*}+U_{2}\delta_{nn'}v_{n,\eta_{p}^{\prime}}v_{n,\eta_{h}^{\prime}}^{*},
\end{equation}
and

\begin{eqnarray}
\langle\text{FS}|c_{\eta_{h}}^{\dagger}c_{\eta_{p}}d_{n}\mathcal{H}d_{n'}^{\dagger}c_{\eta_{p}^{\prime}}^{\dagger}c_{\eta_{h}^{\prime}}|\text{FS}\rangle & = & \left[\delta_{nn'}\left(E_{\text{FS}}+E_{\eta_{p}}-E_{\eta_{h}}\right)+t_{d}\delta_{n\pm1,n'}\right]\delta_{\eta_{p}\eta_{p}^{\prime}}\delta_{\eta_{h}\eta_{h}^{\prime}}\nonumber \\
 &  & +U_{1}\delta_{nn'}\left[\delta_{\eta_{p}\eta_{p}^{\prime}}\delta_{\eta_{h}\eta_{h}^{\prime}}\sum_{E_{\eta}<0}\left|u_{n,\eta}\right|{}^{2}+\delta_{\eta_{h}\eta_{h}^{\prime}}u_{n,\eta_{p}}^{*}u_{n,\eta_{p}^{\prime}}-\delta_{\eta_{p}\eta_{p}^{\prime}}u_{n,\eta_{h}^{\prime}}^{*}u_{n,\eta_{h}}\right]\nonumber \\
 &  & +U_{2}\delta_{nn'}\left[\delta_{\eta_{p}\eta_{p}^{\prime}}\delta_{\eta_{h}\eta_{h}^{\prime}}\sum_{E_{\eta}<0}\left|v_{n,\eta}\right|{}^{2}+\delta_{\eta_{h}\eta_{h}^{\prime}}v_{n,\eta_{p}}^{*}v_{n,\eta_{p}^{\prime}}-\delta_{\eta_{p}\eta_{p}^{\prime}}v_{n,\eta_{h}^{\prime}}^{*}v_{n,\eta_{h}}\right].
\end{eqnarray}
\end{widetext}By using the exact diagonalization technique for a
sparse matrix, we can find the ground polaron state on for the lattice
size $L$ up to 89 (the corresponding dimension of the Hilbert space
is up to several millions) \cite{GreenII}. However, many disorder
realizations are need to address the many-body localization for the
excited polaron states in the middle of the many-body spectrum. For
convenience, we have used the exact diagonalization routine for a
dense matrix and restrict our calculations to $L=13$ (for which the
dimension of the corresponding Hilbert space is $D_{l=6}=2,197$).

\subsection{Quasi-particle properties of the polaron state}

To characterize the quasi-properties of the polaron state, we define
a normalized wave function as: 
\begin{equation}
\begin{aligned}\Psi_{n}=\frac{z_{n}}{\sqrt{\mathcal{Z}}},\end{aligned}
\end{equation}
where $\mathcal{Z}=\sum_{n}z_{n}^{2}$ is the total residue of the
polaron state. Moreover, a useful quantity in characterizing the localization
properties of the ground polaron state is the inverse participation
ratio (IPR). It is given by
\begin{equation}
\begin{aligned}\alpha_{\textrm{IPR}}=\sum_{n=1}^{L}|\Psi_{n}|^{4},\end{aligned}
\end{equation}
which measures the inverse of the number of lattice sites being occupied
by the moving impurity. In the presence of disorder, for a spatially
extended polaron state, it is well known $\alpha_{\textrm{IPR}}\thicksim1/L$,
while for a localized state, $\alpha_{\textrm{IPR}}$ tends to a finite
value at the order of $\mathcal{O}(1)$. The $\alpha_{\textrm{IPR}}$
may also have other size dependence that is different from $1/L$
or $\mathcal{O}(1)$, for possible critical states. By tuning the
disorder disorder strength, we anticipate a sharp change in $\alpha_{\textrm{IPR}}$,
when the system transits from one phase to another. Hence, $\alpha_{\textrm{IPR}}$
can be used to determine the phase boundaries separating the extended,
critical, and localized phases.

\subsection{The MBL indicator}

On the other hand, to investigate the localization properties of high-lying
excited polaron states at infinite temperature or MBL, we adopt a
common quantum chaos indicator, the average ratio between the smallest
and the largest adjacent energy gaps,
\begin{equation}
r_{n}=\frac{\min\left\{ \delta_{n}^{E},\delta_{n-1}^{E}\right\} }{\max\left\{ \delta_{n}^{E},\delta_{n-1}^{E}\right\} },\label{eq: rn}
\end{equation}
where $\delta_{n}^{E}=E_{n}-E_{n-1}$, and $E_{n}$ is the ordered
list of many-body energy levels \cite{VOganesyan2007}. In the thermalized
extended phase, the statistics of the level spacing exhibits a Wigner-Dyson
distribution and the average ratio is $r^{\textrm{WD}}\simeq0.536$,
while in the MBL phase, the statistics of the level spacing exhibits
a Poisson distribution and the average ratio is $r^{\textrm{P}}=2\ln2-1\simeq0.386$
\cite{VOganesyan2007}.

\section{Phase transitions of the ground polaron state }

\label{sec4}

In this section, we discuss the quasi-particle properties of the moving
impurity in its ground state, through the analyses of the wave-function,
inverse participation ratio, energy and residue, as functions of the
disorder strength $V_{0}$, pairing parameter $\Delta$, and atom-impurity
interactions $U_{1}$ and $U_{2}$. This leads to our main result
of the ground-state phase diagram as shown in Fig. \ref{fig1}.

\begin{figure}
\centering{}\includegraphics[width=0.48\textwidth]{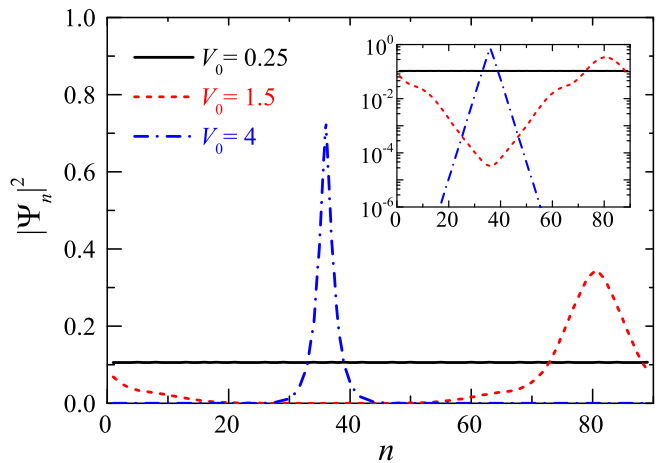} \caption{(color online). The amplitude of the wave function of the ground polaron
state, at three different disorder strengths as indicated. The inset
shows the amplitude in logarithmic scale. Here, we set the length
of the system $L=89$ and the offset phase $\theta=\pi/(2L)$. The
$p$-wave pairing parameter is $\Delta=2t$ and the interaction strengths
between impurity and fermions are $U_{1}=U_{2}=2t$.}
\label{fig2} 
\end{figure}

\subsection{The wave function and inverse participation ratio}

Figure \ref{fig2} reports typical wave-functions of the ground polaron
state in the three different phases for the system length $L=89$,
as we observe with increasing disorder strengths at a nonzero $p$-wave
pairing parameter $\Delta=2t$ and at the atom-impurity interaction
strengths $U_{1}=U_{2}=2t$. At small disorder ($V_{0}=0.25t$, black
solid line), the normalized occupation amplitude of the polaron is
nearly a constant at each site, suggesting that the polaron state
is extended. At a larger disorder strength ($V_{0}=1.5t$, red dashed
line), however, this uniform distribution changes into a broad peak
located near the site $n=80$. Although the occupation amplitude away
from the peak is tiny, it is not exponentially small, as can be seen
from the inset of Fig. \ref{fig2}. It thus seems reasonable to treat
such a state as a \emph{weakly-localized} state. At an even larger
disorder strength ($V_{0}=4t$, blue dot-dashed line), the peak position
moves to the site $n\sim37$ and the width of the peak becomes much
narrower. In addition, away from the peak the occupation amplitude
of the polaron decays exponentially (see the blue dot-dashed line
in the inset), indicating the appearance of a fully or strongly localized
state. 

We remark here that the value of the peak position is of little importance.
In our calculations, a periodic boundary condition is always assumed
and the position of the peak, if exists, is determined by the offset
phase $\theta$, which sets a preferable potential minimum. Therefore,
one can shift the peak position by simply taking a different offset
phase. We also emphasize that the observed localization of the occupation
amplitude of the polaron is induced by interactions between impurity
and fermionic atoms, as the impurity itself does not experience the
quasi-random disorder potential.

\begin{figure}
\centering{}\includegraphics[width=0.48\textwidth]{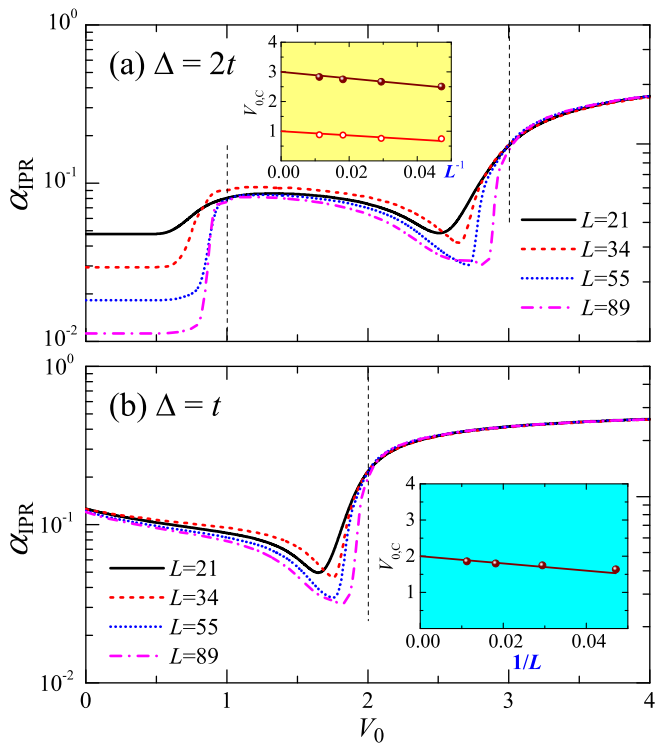} \caption{(color online). The inverse participation ratio $\alpha_{\textrm{IPR}}$
as a function of the quasidisorder strength $V_{0}$ at two $p$-wave
pairing parameters: $\Delta=2t$ (a) and $\Delta=t$ (b). The different
lines correspond to the different sizes of the system, as indicated.
As $V_{0}$ increases, one may identify transitions at some critical
disorder strengths. The two insets reports the length dependence of
the critical strength $V_{0,C}$ and the dashed lines indicate $V_{0,C}$
in the thermodynamic limit $L\rightarrow\infty$. In all the cases,
$U_{1}=U_{2}=2t$ and $\theta=\pi/(2L)$. }
\label{fig3} 
\end{figure}

It is thus clear that with increasing disorder strength, there are
transitions between phases with different localization properties.
As the localization properties of a single-particle state can be conveniently
represented by an inverse participation ratio $\alpha_{\textrm{IPR}}$,
in Fig. \ref{fig3}(a) we report $\alpha_{\textrm{IPR}}$ as a function
of the disorder strength at $\Delta=2t$. Four different system sizes
have been considered, ranging from $L=21$ to $L=89$. It can be seen
that the general behavior of the inverse participation ratio is rather
independent on the system size $L$: $\alpha_{\textrm{IPR}}$ is initially
a constant with increasing disorder strength; At a threshold, it then
jumps suddenly; As the disorder strength increases further, $\alpha_{\textrm{IPR}}$
decreases gradually and exhibits a local minimum before finally rises
and saturates towards a length-independent value. This general behavior
is consistent with the existence of two phase transitions between
three typical wave functions shown in Fig. \ref{fig2}.

\begin{figure}
\centering{}\includegraphics[width=0.4\textwidth]{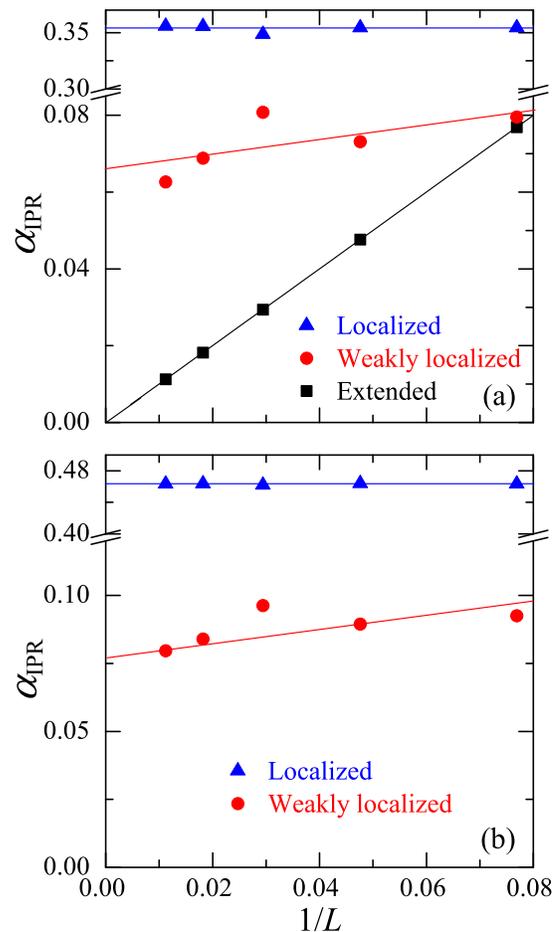}
\caption{(color online). The length dependence of the inverse participation
ratio $\alpha_{\textrm{IPR}}$ in different phases at two $p$-wave
pairing parameters: $\Delta=2t$ (a) and $\Delta=t$ (b). The black
squares, red circles and blue triangles correspond to the selected
cases of extended phase ($V_{0}=0.25t$), weakly localized phase ($V_{0}=1.5t$
in (a) and $V_{0}=t$ in (b)) and strongly localized phase ($V_{0}=4t$
in both (a) and (b)), respectively. For other parameters, we take
$U_{1}=U_{2}=2t$ and $\theta=\pi/(2L)$.}
\label{fig4} 
\end{figure}

For a given $L$, intuitively we may define the inflection point of
the jump and the position of the local minimum as the critical disorder
strengths for the two transitions. In the inset of Fig. \ref{fig3}(a),
we show the critical disorder strengths as a function of the inverse
system size $L^{-1}$. In the thermodynamic limit of $L\rightarrow\infty$,
the critical disorder strengths approach $V_{0,C}=t$ and $V_{0,C}=3t$,
respectively. Interestingly, these two critical disorder strengths
are exactly identical to the two thresholds of the background generalized
AAH model with $p$-wave superfluidity that separate the extended,
critical and localized single-particle states, which are given by
$\left|t-\Delta\right|$ and $t+\Delta$ \cite{JWang2016}, respectively.
By varying the $p$-wave pairing parameter $\Delta$, we have checked
that this identicalness actually holds for any values of $\Delta$.
In Fig. \ref{fig3}(b), we provide another example at $\Delta=t$.
In this case, the first critical disorder strength decreases to $V_{0,C}=\left|t-\Delta\right|=0$
and therefore cannot be identified from the plot. 

As a brief conclusion, we find that the moving impurity or polaron
in the ground state shares a similar phase diagram as the background
fermionic atoms, which has already been illustrated by the figure
of $\alpha_{\textrm{IPR}}$ in logarithmic scale for a small system
size $L=21$, as shown in Fig. \ref{fig1}. This is a very useful
observation, since it is then reasonable to anticipate that the measurements
of the polaron quasi-particle properties, such as its energy and residue,
could provide a useful probe of the background non-abelian AAH model
with $p$-wave superfluidity. 

\begin{figure}
\centering{}\includegraphics[width=0.4\textwidth]{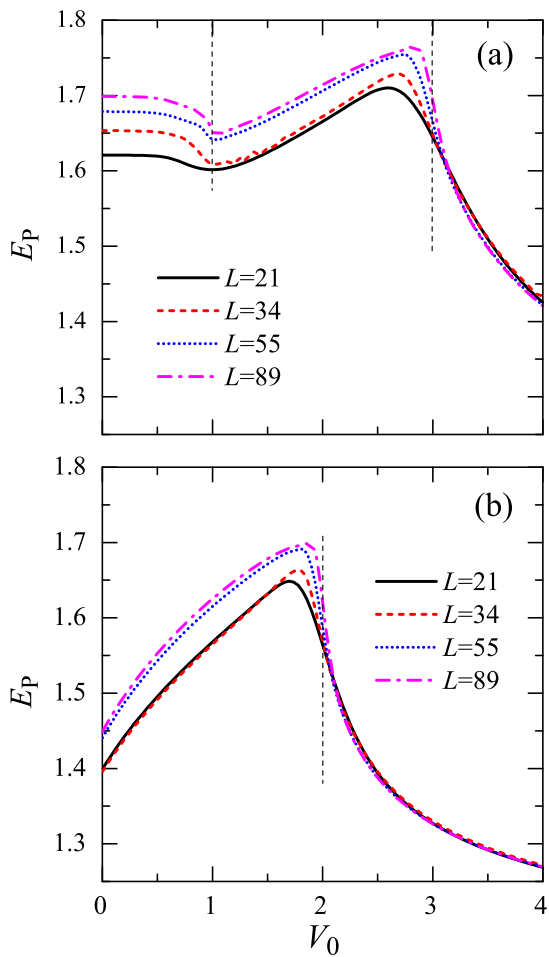}
\caption{(color online). The polaron energy $E_{P}$ as a function of the quasidisorder
strength $V_{0}$ at two $p$-wave pairing parameters: $\Delta=2t$
(a) and $\Delta=t$ (b). The different lines correspond to the different
sizes of the system, as indicated. Here, we take $U_{1}=U_{2}=2t$
and $\theta=\pi/(2L)$. }
\label{fig5} 
\end{figure}

The only difference between the two phase diagrams is that the critical
phase in the non-abelian AAH model, enclosed by the curves $\left|t-\Delta\right|$
and $t+\Delta$, has now been replaced by the weakly localized phase
of the polaron. This is not surprising, since in some sense the critical
phase of fermionic atoms is fragile and may not be mirrored by the
polaron. To confirm it, we check the inverse participation ratio $\alpha_{\textrm{IPR}}$
of the three typical wave-functions in the thermodynamic limit $L\rightarrow\infty$.
As shown in Fig. \ref{fig4}(a), $\alpha_{\textrm{IPR}}$ of the extended
state (black squares) vanishes linearly as a function of $1/L$. On
the contrary, $\alpha_{\textrm{IPR}}$ of the localized state (blue
triangles) takes a finite value at the order of $\mathcal{O}(1)$
and is essentially unchanged with decreasing $1/L$. The intermediate
state (red circles) seems to have much smaller $\alpha_{\textrm{IPR}}$
than the localized state. But, it does not vanish as $L^{-1}\rightarrow0$,
unlike a critical phase. This justifies the use of our terminology
of a weakly localized state.

\begin{figure}
\centering{}\includegraphics[width=0.4\textwidth]{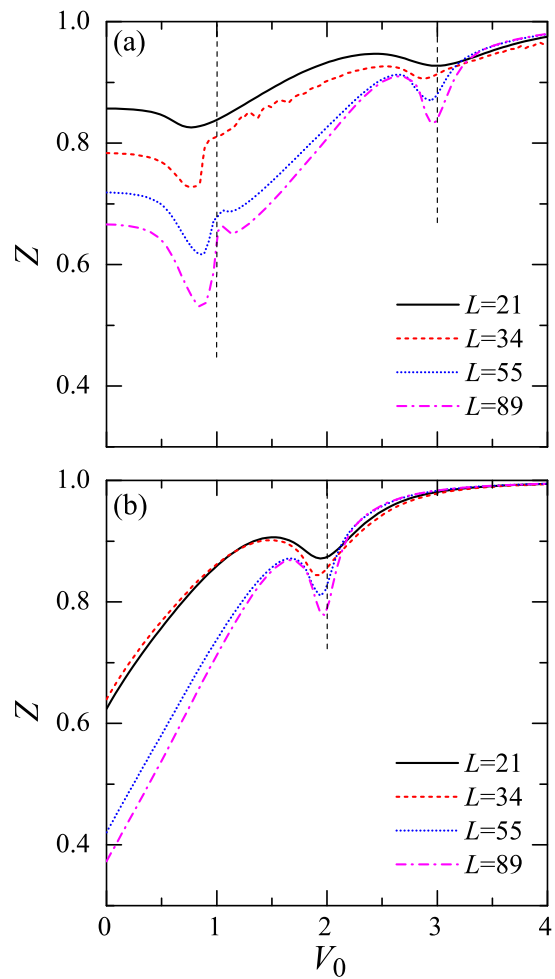}
\caption{(color online). The polaron residue $\mathcal{Z}$ as a function of
the quasidisorder strength $V_{0}$ at two $p$-wave pairing parameters:
$\Delta=2t$ (a) and $\Delta=t$ (b). The different lines correspond
to the different sizes of the system, as indicated. Here, we take
$U_{1}=U_{2}=2t$ and $\theta=\pi/(2L)$. }
\label{fig6} 
\end{figure}

\begin{figure}
\centering{}{\LARGE{}\includegraphics[width=0.45\textwidth]{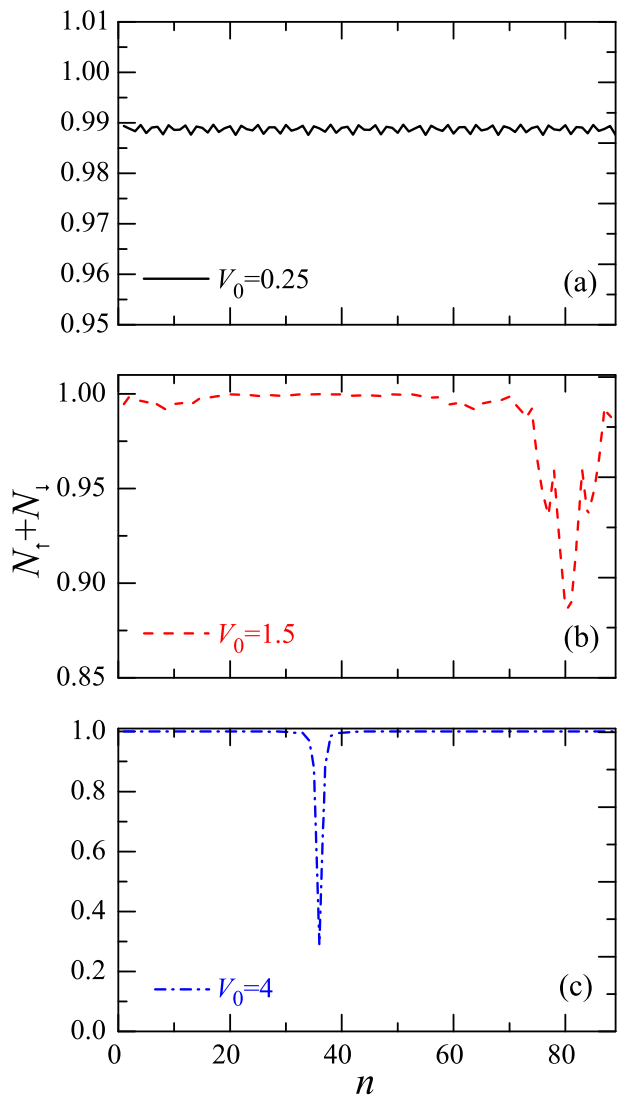}}
\caption{(color online). The occupation number of fermionic atoms at three
different disorder strengths: (a) $V_{0}=0.25t$, (b) $V_{0}=1.5t$,
and (c) $V_{0}=4t$. Here, we set the length of the system $L=89$
and the offset phase $\theta=\pi/(2L)$. The $p$-wave pairing parameter
is $\Delta=2t$ and the interaction strengths between impurity and
fermions are $U_{1}=U_{2}=2t$.}
\label{fig7} 
\end{figure}

\subsection{The polaron energy and residue}

Fig. \ref{fig5} and Fig. \ref{fig6} present the polaron energy $E_{P}=E-E_{\textrm{FS}}-E_{\textrm{imp}}^{(0)}$
and the polaron residue $\mathcal{Z}$ as a function of the disorder
strength at $U_{1}=U_{2}=2t$, respectively. Here, $E$ is the energy
of the ground state obtained by exact diagonalization and $E_{\textrm{imp}}^{(0)}=-2t_{d}=-2t$.
As anticipated, both energy and residue show non-monotonic dependences
on the disorder strength, which are consistent with the existence
of some phase transitions. Let us focus on the energy at $\Delta=2t$,
as shown in Fig. \ref{fig5}(a). At small disorder, the energy of
the polaron slowly decreases with increasing disorder strength. This
is because in the extended state, the fermionic atoms and impurity
are miscible and are able to optimize their distance to reduce the
repulsive interaction energy. At large disorder, the energy again
decreases as the disorder strength increases. In this limit, the atoms
and impurity are essentially phase separated. This is evident by comparing
the blue dot-dashed lines in Fig. \ref{fig7}(c) and in Fig. \ref{fig2},
which show the atomic density distribution and the impurity density
distribution, respectively. The phase separation favors a small repulsive
interaction energy, which in turns provides a mechanism for the full
localization of the polaron. At an intermediate disorder strength
(i.e., $t<V_{0}<3t$ in Fig. \ref{fig5}(a)), the system is actually
frustrated. The fermionic atoms and impurity try to avoid each other
to reduce the interaction energy, but the strength of the disorder
is not large enough to create a well-localized polaron state. This
leads to a slight oscillation of the atomic density distribution around
the impurity, as can be seen from Fig. \ref{fig7}(b). The frustration
is responsible for the enhancement of the polaron energy shown in
\ref{fig5}(a), as the disorder strength increases.

\begin{figure}
\centering{}\includegraphics[width=0.48\textwidth]{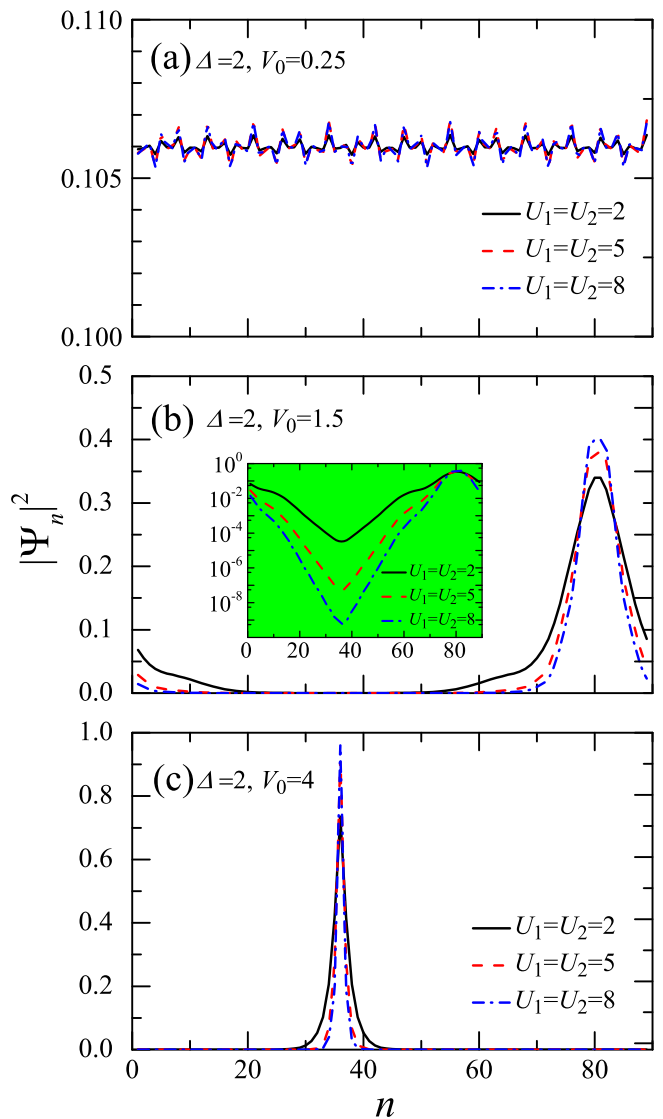}
\caption{(color online). The amplitude of the wave function of the ground polaron
state at three different disorder strengths: (a) $V_{0}=0.25t$, (b)
$V_{0}=1.5t$, and (c) $V_{0}=4t$ and at three different interaction
strengths as indicated. The inset in (b) shows the amplitude in logarithmic
scale. Here, we set the length of the system $L=89$ and the offset
phase $\theta=\pi/(2L)$. The $p$-wave pairing parameter is $\Delta=2t$.}
\label{fig8} 
\end{figure}

\subsection{The dependence on the atom-impurity interaction}

We consider so far for the case of fixed atom-impurity interaction
strengths $U_{1}=U_{2}=2t$. Nevertheless, the phase diagram of the
ground polaron state does not change if we take larger interaction
strength. In Fig. \ref{fig8}, we report the three typical wave-functions
at different interaction strengths. In the extended phase (a), the
wave-function is basically unchanged upon increasing interaction strength.
On the other hand, in the weakly localized phase (b) or the localized
phase (c), the interaction strength tends to sharpen the localization
peak and hence make the state more localized. One may expect that
the weakly localized phase turns into a well localized phase at sufficiently
large interaction strength. This seems unlikely, however, since the
full width at half maximum (FWHM) of the peak of the weakly localized
phase does not decrease too much with increasing interaction strength.
It remains large at $U_{1}=U_{2}=8t$ (i.e., $\sim10$ sites), much
larger than that of a fully localized phase (i.e., $\sim2$ sites)

\begin{figure}
\centering{}{\LARGE{}\includegraphics[width=0.48\textwidth]{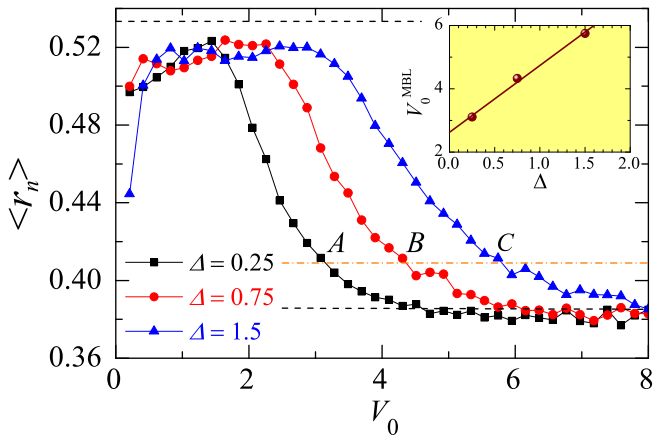}}
\caption{(color online). The averaged ratio of adjacent energy gaps $\langle r_{n}\rangle$
as a function of the disorder strength at different values of the
$p$-wave pairing parameter $\Delta$. The average is calculated over
the central half of the spectrum, averaging over 100 quasi-random
disorder realizations, by randomly generating the phase offset $\theta$.
The two dashed lines show the anticipated average ratios for the Wigner-Dyson
distribution ($r^{\textrm{WD}}=0.536$) and for the Poisson distribution
($r^{\textrm{P}}=0.386$), respectively. The dot-dashed line indicates
a critical averaged ratio $\langle r_{n}\rangle_{\textrm{MBL}}=0.41$,
below which all the states of the polaron may become localized, i.e.,
MBL occurs. The inset shows the critical disorder strength for MBL
as a function of the pairing parameter $\Delta$, with a straight
line as a guide to the eye. Here, we take $U_{1}=U_{2}=2$ and a small
system size $L=13$.}
\label{fig9} 
\end{figure}

\section{many-body localization of the excited polaron states }

\label{sec5}

We now turn to consider the localization of the high-energy polaron
states or MBL. A convenient way to identify the MBL is to calculate
the averaged ratio of adjacent energy levels, which is defined in
Eq. (\ref{eq: rn}). In Fig. \ref{fig9}, we report the averaged ratio
$\langle r_{n}\rangle$ as a function of the disorder strength for
different values of the $p$-wave pairing parameter $\Delta$. Here,
we take $U_{1}=U_{2}=2$ and $L=13$. The three different lines correspond
to the different values of $\Delta$. The two horizontal dashed lines
show the averaged ratio for the Wigner-Dyson distribution ($r^{\textrm{WD}}\simeq0.536$)
and for the Poisson distribution ($r^{\textrm{P}}\simeq0.386$), respectively.
Quite generally, we find that when the disorder is weak, $\langle r_{n}\rangle$
tends to $r^{\textrm{WD}}$, which means the system can be thermalized.
In contrast, when the disorder is strong, $\langle r_{n}\rangle$
approaches $r^{\textrm{P}}$, indicating the appearance of MBL.

It is readily seen that the averaged ratio depends sensitively on
the pairing parameter. By increasing $\Delta$, the curve is shifted
horizontally to the right side of the figure. The system thus seems
to become more difficult to be many-body localized as $\Delta$ increases.
To have a qualitative characterization, we may estimate the critical
disorder strength of MBL, $V_{0}^{\textrm{MBL}}$, by using the criterion,
\begin{equation}
\left\langle r_{n}\right\rangle (V_{0}^{\textrm{MBL}})=0.41.
\end{equation}
This naïve estimation is qualitative only and is motivated by the
fact that the MBL in the disordered spin-chain \cite{VOganesyan2007,DJLuitz2015}
and Hubbard models \cite{RMondaini2015} occurs at a similar averaged
ratio $\left\langle r_{n}\right\rangle _{\textrm{MBL}}\sim0.41$.
In the figure, we show this MBL averaged ratio by a dot-dashed line
and determine $V_{0}^{\textrm{MBL}}$ at the three pairing parameters
from the three cross points, labelled as $A$, $B$ and $C$, respectively.
As shown in the inset, roughly speaking, the critical disorder strength
for MBL $V_{0}^{\textrm{MBL}}$ depends linearly on the $p$-wave
pairing parameter $\Delta$.

\section{Conclusions}

\label{sec6}

In summary, we have theoretically investigated the ground-state and
excited states properties of a mobile impurity or polaron immersed
in a one-dimensional quasiperiodic Fermi system with topological $p$-wave
superfluidity. On the one hand, for the ground state we find four
distinct phases of the polaron: two extend phases, a weakly localized
phase and a localized phase (see Fig. \ref{fig1}), according to the
inverse participation ratio of the polaron wave-function. This phase
diagram of the polaron is a perfect mirror of the phase diagram of
the background fermionic atoms, which is described by the non-abelian
Aubry-André-Harper model. As a result, experimentally we may probe
the phase diagram of the non-abelian AAH model by measuring the quasi-particle
properties of the polaron. On the other hand, we have briefly considered
the many-body localization of the excited polaron states at infinite
temperature. We find that the existence of a $p$-wave pairing parameter
helps delocalize MBL and makes the system easier to thermalize.
\begin{acknowledgments}
This work was supported by Australian Research Council Future Fellowship
grants (Grants No. FT140100003 and No. FT130100815) and Discovery
Projects (Grants No. DP170104008 and No. DE180100592). J. Xiong was
supported by the National Natural Science Foundation of China under
Grant No. 11474027. F.-G. Deng was supported by the National Natural
Science Foundation of China under Grant No. 11474026 and No. 11674033,
and the Fundamental Research Funds for the Central Universities under
Grant No. 2015KJJCA01. This work was performed on the swinSTAR supercomputer
at Swinburne University of Technology. 
\end{acknowledgments}


\begin{thebibliography}{10}
\bibitem{PWAnderson1958} P. W. Anderson, Phys. Rev. \textbf{109},
1492 (1958).

\bibitem{RDalichaouch1991} R. Dalichaouch, J. P. Armstrong, S. Schultz,
P. M. Platzman, and S. L. McCall, Nature (London) \textbf{354}, 53
(1991). 

\bibitem{AAChabanov2000} A. A. Chabanov, M. Stoytchev, and A. Z.
Genack, Nature (London) \textbf{404}, 850 (2000).

\bibitem{TSchwartz2007} T. Schwartz, G. Bartal, S. Fishman, and M.
Segev, Nature (London) \textbf{446}, 52 (2007). 

\bibitem{YLahini2008} Y. Lahini, A. Avidan, F. Pozzi, M. Sorel, R.
Morandotti, D. N. Christodoulides, and Y. Silberberg, Phys. Rev. Lett.
\textbf{100}, 013906 (2008).

\bibitem{JBilly2008} J. Billy, V. Josse, Z. Zuo, A. Bernard, B. Hambrecht,
P. Lugan, D. Clement, L. Sanchez-Palencia, P. Bouyer, and A. Aspect,
Nature (London) \textbf{453}, 891 (2008); 

\bibitem{GRoati2008} G. Roati, C. D'Errico, L. Fallani, M. Fattori,
C. Fort, M. Zaccanti, G. Modugno, M. Modugno, and M. Inguscio, Nature
(London) \textbf{453}, 895 (2008).

\bibitem{MRigol2008} M. Rigol, V. Dunjko, and M. Olshanii, Nature
(London) \textbf{452}, 854 (2008). 

\bibitem{APolkovnikov2011} A. Polkovnikov, K. Sengupta, A. Silva,
and M. Vengalattore, Rev. Mod. Phys. \textbf{83}, 863 (2011).

\bibitem{JMDeutsch1991} J. M. Deutsch, Phys. Rev. A \textbf{43},
2046 (1991). 

\bibitem{MSrednicki1994} M. Srednicki, Phys. Rev. E \textbf{50},
888 (1994).

\bibitem{RNandkishore2015} R. Nandkishore and D. A. Huse, Annu. Rev.
Condens. Matter Phys. \textbf{6}, 15 (2015). 

\bibitem{EAltman2015} E. Altman and R. Vosk, Annu. Rev. Condens.
Matter Phys. \textbf{6}, 383 (2015).

\bibitem{APal2010} A. Pal and D. A. Huse, Phys. Rev. B \textbf{82},
174411 (2010).

\bibitem{JHBardarson2012} J. H. Bardarson, F. Pollmann, and J. E.
Moore, Phys. Rev. Lett. \textbf{109}, 017202 (2012). 

\bibitem{RVosk2013} R. Vosk and E. Altman, Phys. Rev. Lett. \textbf{110},
067204 (2013). 

\bibitem{MSerbyn2013} M. Serbyn, Z. Papic, and D. A. Abanin, Phys.
Rev. Lett. \textbf{110}, 260601 (2013). 

\bibitem{FAndraschko2014} F. Andraschko, T. Enss, and J. Sirker,
Phys. Rev. Lett. \textbf{113}, 217201 (2014).

\bibitem{MSchreiber2015} M. Schreiber, S. S. Hodgman, P. Bordia,
H. P. Lüschen, M. H. Fischer, R. Vosk, E. Altman, U. Schneider, and
I. Bloch, Science \textbf{349}, 842 (2015).

\bibitem{PGHarper1955} P. G. Harper, Proc. Phys. Soc. London, Sect.
A \textbf{68}, 874 (1955). 

\bibitem{SAubry1980} S. Aubry and G. André, Ann. Israel Phys. Soc.
\textbf{3}, 133 (1980).

\bibitem{SIyer2013} S. Iyer, V. Oganesyan, G. Refael, and D. A. Huse,
Phys. Rev. B \textbf{87}, 134202 (2013).

\bibitem{SOstlund1983} S. Ostlund, R. Pandit, D. Rand, H. J. Schellnhuber,
and E. D. Siggia, Phys. Rev. Lett. \textbf{50}, 1873 (1983). 

\bibitem{MKohmoto1983} M. Kohmoto, Phys. Rev. Lett. \textbf{51},
1198 (1983). 

\bibitem{DJThouless1983} D. J. Thouless, Phys. Rev. B \textbf{28},
4272 (1983). 

\bibitem{JHHan1994} J. H. Han, D. J. Thouless, H. Hiramoto, and M.
Kohmoto, Phys. Rev. B \textbf{50}, 11365 (1994). 

\bibitem{IChang1997} I. Chang, K. Ikezawa, and M. Kohmoto, Phys.
Rev. B \textbf{55}, 12971 (1997). 

\bibitem{YTakada2004} Y. Takada, K. Ino, and M. Yamanaka, Phys. Rev.
E \textbf{70}, 066203 (2004). 

\bibitem{FLiu2015} F. Liu, S. Ghosh, and Y. D. Chong, Phys. Rev.
B \textbf{91}, 014108 (2015).

\bibitem{LJLang2012} L.-J. Lang, X. Cai, and S. Chen, Phys. Rev.
Lett. \textbf{108}, 220401 (2012). 

\bibitem{YEKraus2012} Y. E. Kraus and O. Zilberberg, Phys. Rev. Lett.
\textbf{109}, 116404 (2012). 

\bibitem{SLZhu2013} S.-L. Zhu, Z.-D.Wang, Y.-H. Chan, and L.-M. Duan,
Phys. Rev. Lett. \textbf{110}, 075303 (2013). 

\bibitem{WDeGottardi2013} W. DeGottardi, D. Sen, and S. Vishveshwara,
Phys. Rev. Lett. \textbf{110}, 146404 (2013).

\bibitem{XCai2013} X. Cai, L.-J. Lang, S. Chen, and Y. Wang, Phys.
Rev. Lett. \textbf{110}, 176403 (2013).

\bibitem{FGrusdt2013} F. Grusdt, M. Honing, and M. Fleischhauer,
Phys. Rev. Lett. \textbf{110}, 260405 (2013). 

\bibitem{IISatija2013} I. I. Satija and G. G. Naumis, Phys. Rev.
B \textbf{88}, 054204 (2013). 

\bibitem{RBarnett2013} R. Barnett, Phys. Rev. A \textbf{88}, 063631
(2013). 

\bibitem{XDeng2014} X. Deng and L. Santos, Phys. Rev. A \textbf{89},
033632 (2014).

\bibitem{JWang2016} J. Wang, X.-J. Liu, G. Xianlong, and H. Hu, Phys.
Rev. B \textbf{93}, 104504 (2016).

\bibitem{Zeng2017}Q.-B. Zeng, S. Chen, and R. Lü, Phys. Rev. A \textbf{95},
062118 (2017). 

\bibitem{FChevy2010}F. Chevy and C. Mora, Rep. Prog. Phys. \textbf{73},
112401 (2010).

\bibitem{PMassignan2014}P. Massignan, M. Zaccanti, and G. M. Bruun,
Rep. Prog. Phys. \textbf{77}, 034401 (2014).

\bibitem{AVBalatsky2006}A. V. Balatsky, I. Vekhter, and J.-X. Zhu,
Rev. Mod. Phys. \textbf{78}, 373 (2006). 

\bibitem{EWHudson2001}E. W. Hudson, K. M. Lang, V. Madhavan, S. H.
Pan, H. Eisaki, S. Uchida, and J. C. Davis, Nature (London) \textbf{411},
920 (2001).

\bibitem{HHu2013}H. Hu, L. Jiang, H. Pu, Y. Chen, and X.-J. Liu,
Phys. Rev. Lett. \textbf{110}, 020401 (2013).

\bibitem{HHu2016}H. Hu, A. B. Wang, S. Yi, and X.-J. Liu, Phys. Rev.
A \textbf{93}, 053601 (2016).

\bibitem{FChevy2006}F. Chevy, Phys. Rev. A \textbf{74}, 063628 (2006).

\bibitem{GreenII}For more details on Swinburne supercomputer Green
II resource, we refer to the homepage, http://supercomputing.swin.edu.au/.

\bibitem{VOganesyan2007} V. Oganesyan and D. A. Huse, Phys. Rev.
B \textbf{75}, 155111 (2007).

\bibitem{DJLuitz2015} D. J. Luitz, N. Laflorencie, and F. Alet, Phys.
Rev. B \textbf{91}, 081103(R) (2015).

\bibitem{RMondaini2015} R. Mondaini and M. Rigol, Phys. Rev. A \textbf{92},
041601(R) (2015). 
\end{thebibliography}
\end{document}